\documentclass[aps,amsmath,twocolumn,amssymb,floatfixng,showpacs,
superscriptaddress]{revtex4}

\usepackage[dvips]{graphics}
\usepackage{color}
\definecolor{gold}{rgb}{0.85,0.66,0}
\definecolor{dred}{rgb}{0.6,0,0}

\begin{document}

\title{Magnetic Response in a Zigzag Carbon Nanotube}

\author{Paramita Dutta}

\affiliation{Theoretical Condensed Matter Physics Division, Saha 
Institute of Nuclear Physics, Sector-I, Block-AF, Bidhannagar, 
Kolkata-700 064, India} 

\author{Santanu K. Maiti}

\email{santanu@post.tau.ac.il}

\affiliation{School of Chemistry, Tel Aviv University, Ramat-Aviv,
Tel Aviv-69978, Israel}

\author{S. N. Karmakar}

\affiliation{Theoretical Condensed Matter Physics Division, Saha 
Institute of Nuclear Physics, Sector-I, Block-AF, Bidhannagar, 
Kolkata-700 064, India} 

\begin{abstract}
Magnetic response of interacting electrons in a zigzag carbon nanotube 
threaded by a magnetic flux is investigated within a Hartree-Fock mean 
field approach. Following the description of energy spectra for both
non-interacting and interacting cases we analyze the behavior of
persistent current in individual branches of a nanotube. Our present
investigation leads to a possibility of getting a filling-dependent 
metal-insulator transition in a zigzag carbon nanotube.
\end{abstract}


\maketitle

\section{Introduction}

The isolation of single layer graphene by Novoselov {\it et al.}~\cite{geim1}
has initiated intense and diverse research on this system. Graphene, a 
single layer of carbon atoms tightly packed into a two-dimensional 
honey-comb lattice, has drawn attention of scientists in various 
disciplines due to its unconventional and fascinating electronic 
properties arising particularly from the linear dispersion relation 
around the Dirac points of the hexagonal Brillouin zone. These unique 
properties can be understood in terms of the Dirac Hamiltonian~\cite{neto} 
since it actually describes the physics of electrons near the Fermi level 
of the undoped material. The carriers in graphene effectively behave as 
massless relativistic particles within a low energy range close to Fermi 
energy and these massless Dirac Fermions~\cite{geim2} evince various 
phenomena in this energy range. The bipartite character of the wonderful 
lattice structure of graphene strongly 
influences its intrinsic properties and makes graphene a wonderful testbed 
not only for condensed matter theory, but also for quantum field theory 
and mathematical physics. Though lot of studies have been done both
theoretically as well as experimentally to reveal electronic properties
of this exotic system, yet complete knowledge about it is still lacking.
This motivates us to address some interesting issues of electron transport
in carbon nanotubes where a nanotube is formed by rolling up a graphite
ribbon in the cylindrical form~\cite{kit}. 

In this article we explore the behavior of persistent current in a finite 
sized graphite nanotube with zigzag edges within a nearest-neighbor 
tight-binding (TB) framework using a generalized Hartree-Fock (HF) 
approximation~\cite{san,san0,fractal}. The phenomenon of persistent current 
in a mesoscopic ring, threaded by an Aharonov-Bohm (AB) flux $\phi$, has 
been studied theoretically more than two decades ago~\cite{butt,gefen1,
schm,ambe}. Later it has been justified through several nice 
experiments~\cite{levy,chand,jari,deb}. It is a pure quantum mechanical 
effect and can persist without any dissipation in the equilibrium case. 
In the last few years extensive studies on persistent current in carbon 
nanotubes have been performed and many interesting physical phenomena have 
been explored~\cite{sasaki,chen,szopa}. Persistent current in a carbon 
nanotube is highly sensitive to its radius, chirality, deformation, etc. 
Very recently it has also been observed experimentally that the Fermi 
energy of a carbon nanotube can be regulated nicely by means of electron 
or hole doping, which can induce a dramatic change in persistent 
current~\cite{szopa}. It is well established that in a conventional 
multi-channel mesoscopic cylinder electron transport strongly depends on 
the correlation among different channels as well as the shape of
\begin{figure}[ht]
{\centering \resizebox*{7cm}{5cm}{\includegraphics{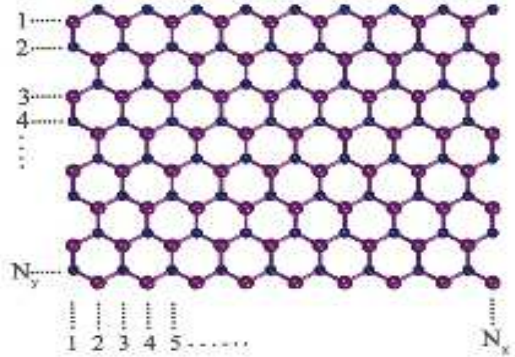}}\par}
\caption{(Color online). Schematic view of a zigzag graphite nano-ribbon 
with $N_x$ and $N_y$ number of atomic sites along the $x$ and $y$ 
directions, respectively.}
\label{zig}
\end{figure}
Fermi surface. Therefore we might expect some interesting features
of persistent current in a carbon nanotube due to its unique
electronic structure.

The behavior of persistent current in zigzag nanotubes has been addressed 
theoretically by some groups~\cite{sasaki,chen,szopa}. There are many 
theoretical techniques available in the literature~\cite{gefen2,san1,san2,
belu,ore,peeters} which generally investigate magnetic response of the 
entire system but the distribution of persistent current among different 
branches of the nanotube remains unaddressed though it is highly important 
to illuminate the magnetic response of the system with a deeper insight. 
To the best of our knowledge, no rigorous effort has been made so far 
to unravel the behavior of persistent current in separate branches of 
a graphite nanotube. This is the main motivation behind this work.

In what follows, we present the results. Section II is devoted to present 
the model and generalized HF approach. Following the energy spectra for 
the non-interacting and interacting cases (Section III), in Section IV we 
establish the second quantized form to evaluate persistent current in 
individual branches of a zigzag carbon nanotube. The energy-flux 
characteristics are described in Section V, while the behavior of 
persistent current in separate branches of a nanotube is illustrated in 
Section VI. Finally, in Section VII we draw our conclusions.

\section{The Model and the Mean Field Approach}

\begin{figure}[ht]
{\centering \resizebox*{3cm}{3cm}{\includegraphics{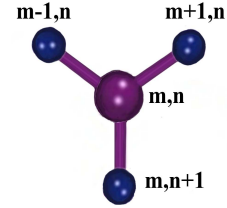}}\par}
\caption{(Color online). Schematic view of different atomic sites
with their co-ordinates.}
\label{index}
\end{figure}

We begin with a graphite nano-ribbon of zigzag edges as shown in 
Fig.~\ref{zig}, where the filled magenta (large) and blue (small) circles 
correspond to two different sub-lattices, namely, A and B, respectively. 
$N_x$ and $N_y$ correspond to the number of atomic sites 
\begin{figure}[ht]
{\centering \resizebox*{3cm}{4.5cm}{\includegraphics{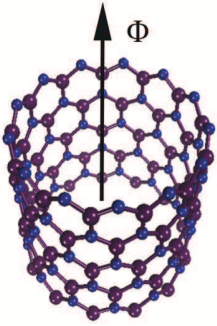}}\par}
\caption{(Color online). A graphite nanotube threaded by a magnetic flux 
$\Phi$.}
\label{tube}
\end{figure}
along the $x$ and $y$ directions, respectively. In order to elucidate 
magnetic response of a nanotube we roll up the graphite ribbon along 
$x$ direction using periodic boundary condition and allow to pass a 
magnetic flux $\phi$ (measured in unit of elementary flux quantum 
$\phi_0=c h/e$) along the axis of the tube as shown in Fig.~\ref{tube}. 
We describe our model quantum system by the nearest-neighbor TB framework 
which captures most of the essential properties of the tube 
nicely~\cite{lopez,lin,sorella,heyd}. In presence of magnetic flux 
$\phi$, the Hamiltonian of an interacting zigzag nanotube reads,
\begin{eqnarray}
H & = & t \sum_{m,n,\sigma} \left(a_{m,n,\sigma}^{\dag}b_{m-1,n,\sigma}
e^{-i \theta} + a_{m,n,\sigma}^{\dag} b_{m+1,n,\sigma}\,e^{i \theta} \right. 
\nonumber \\
& + & \left. a_{m,n,\sigma}^{\dag} b_{m,n+1,\sigma}\right) + \mbox{h.c.} 
\nonumber \\
& +& U \sum_{m,n} \left(a_{m,n,\uparrow}^{\dag}
a_{m,n\uparrow}a_{m,n,\downarrow}^{\dag}a_{m,n,\downarrow} \right. \nonumber \\
&+&\left. b_{m+1,n,\uparrow}^{\dag}b_{m+1,n,\uparrow}
b_{m+1,n,\downarrow}^{\dag}b_{m+1,n,\downarrow} \right) 
\label{ham}
\end{eqnarray}
where, $m$ and $n$ are integers describing the co-ordinates of the lattice
sites. The site indexing is schematically shown in Fig.~\ref{index} for 
better viewing. $t$ is the nearest-neighbor hopping integral, 
$a_{m,n}^{\dag}$ ($b_{m,n}^{\dag}$) is the creation operator for an 
electron of spin $\sigma$ ($\uparrow$,$\downarrow$) associated with A (B) 
type of 
sites at the position ($m$,$n$) and the corresponding annihilation operator 
is denoted by $a_{m,n}$ ($b_{m,n}$). The factor $\theta$ ($=2 \pi \phi/N_x$), 
the so-called Peierl's phase factor, is introduced into the above Hamiltonian 
to incorporate the effect of magnetic flux applied along the axis of the 
tube. $U$ is the strength of on-site Hubbard interaction.
\vskip 0.2cm
\noindent
{\em \underline{Decoupling of interacting Hamiltonian}:} Using the 
generalized HF approach, we decouple the TB Hamiltonian into two different 
parts corresponding to two different values of $\sigma$ ($\uparrow$ and 
$\downarrow$). After decoupling, the Hamiltonian looks like,
\begin{equation}
H_{\mbox{\tiny MF}}=H_{\uparrow}+H_{\downarrow}+H_0
\end{equation}
where,
\begin{eqnarray}
H_{\uparrow} & = & U \sum_{m,n} \left(\langle n^a_{m,n,\downarrow}\rangle
n^a_{m,n,\uparrow} + \langle n^b_{m+1,n,\downarrow}\rangle n^b_{m+1,n,
\uparrow} \right) \nonumber \\
& + & t \sum_{m,n} \left(a_{m,n,\uparrow}^{\dag}b_{m-1,n,\uparrow}
e^{-i \theta} + a_{m,n,\uparrow}^{\dag} b_{m+1,n,\uparrow}\,e^{i \theta} 
\right.  \nonumber \\
& + & \left. a_{m,n,\uparrow}^{\dag} b_{m,n+1,\uparrow} + \mbox{h.c.} 
\right),
\end{eqnarray}
\begin{eqnarray}
H_{\downarrow}&=& U \sum_{m,n} \left(\langle n^a_{m,n,\uparrow}\rangle 
n^a_{m,n,\downarrow} + \langle n^b_{m+1,n,\uparrow}\rangle 
n^b_{m+1,n,\downarrow}\right) \nonumber \\
& + & t \sum_{m,n} \left(a_{m,n,\downarrow}^{\dag}
b_{m-1,n,\downarrow}e^{-i \theta} + a_{m,n,\downarrow}^{\dag} 
b_{m+1,n,\downarrow}\,e^{i \theta} \right. \nonumber \\
& + & \left. a_{m,n,\downarrow}^{\dag} b_{m,n+1,\downarrow}+ 
\mbox{h.c.} \right)
\end{eqnarray}
and
\begin{eqnarray}
H_0 & = & - U \sum_{m,n}\left(\langle n^a_{m,n,\uparrow}\rangle 
\langle n^a_{m,n,\downarrow}\rangle \right. \nonumber \\ 
& + & \left. \langle n^b_{m+1,n,\uparrow}\rangle 
\langle n^b_{m+1,n,\downarrow}\rangle \right).
\end{eqnarray}
Here, $n^a_{m,n,\sigma}$ and $n^b_{m,n,\sigma}$ are the number operators 
associated with the A and B types of atoms, respectively. $H_{\uparrow}$ 
and $H_{\downarrow}$ are the Hamiltonians for up and down spin electrons, 
respectively. $H_0$ is a constant term which gives the energy shift.
\vskip 0.2cm
\noindent
{\em \underline{Self-consistent procedure}:}
In order to get the energy eigenvalues of the interacting Hamiltonian we go 
through a self-consistent procedure considering initial guess values of 
$\langle n^a_{m,n,\sigma}\rangle$ and $\langle n^b_{m,n,\sigma}\rangle$. 
With these initial values, the up and down spin Hamiltonians are 
diagonalized numerically and a new set of values of 
$\langle n^a_{m,n,\sigma}\rangle$ and $\langle n^b_{m,n,\sigma}\rangle$ 
are calculated. These steps are repeated until a self-consistent solution 
is achieved. 
\vskip 0.2cm
\noindent
{\em \underline{Finding the ground state energy}:}
After getting the self-consistent solution we determine the ground state 
energy ($E_0$) at absolute zero temperature ($T=0\,K$) for a particular 
filling by taking the sum of individual states upto the Fermi level 
($E_F$) for both up and down spin electrons. The expression for ground 
state energy reads,
\begin{equation}
E_0=\sum_{i}E_{i,\uparrow}+\sum_{i}E_{i,\downarrow}+H_0
\end{equation} 
where, $i$ runs over the states up to the Fermi level. $E_{i,\uparrow}$'s 
and $E_{i,\downarrow}$'s are the single particle energy eigenvalues obtained 
by diagonalizing the up and down spin Hamiltonians $H_{\uparrow}$ and 
$H_{\downarrow}$, respectively.

\section{Energy spectrum}

To make this present communication a self contained study let us first 
start with the energy band structure of a finite width zigzag nano-ribbon. 
\vskip 0.2cm
\noindent
{\em \underline{Non-interacting case}:}
To establish the energy dispersion relation of a zigzag nano-ribbon
\begin{figure}[ht]
{\centering \resizebox*{3.5cm}{5cm}{\includegraphics{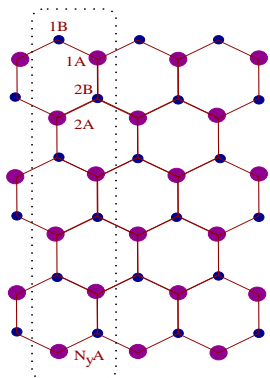}}\par}
\caption{(Color online). Unit cell configuration of a 
zigzag nano-ribbon.}
\label{unit}
\end{figure}
we find an effective difference equation analogous to the case of an 
infinite one-dimensional chain.
This can be done by proper choice of a unit cell from the nano-ribbon. 
The schematic view of a unit cell configuration with $N_y$ pairs of B-A 
atoms in a zigzag nano-ribbon is shown in Fig.~\ref{unit}. With this 
arrangement, the effective difference equation of the nano-ribbon gets 
the form,
\begin{equation}
(E \,\mathcal{I} - \mathcal{E}_{\sigma})\psi_{j,\sigma} = \mathcal{T} 
\psi_{j+1,\sigma} + \mathcal{T}^{\dag} \psi_{j-1,\sigma} 
\label{diff}
\end{equation}
where, 
\begin{eqnarray}
\psi_{j,\sigma}= \left(\begin{array}{c}
\psi_{j1B,\sigma} \\
\psi_{j1A,\sigma} \\
\psi_{j2B,\sigma} \\
.\\
.\\
\psi_{jN_yA,\sigma}\end{array}\right).
\end{eqnarray}
$\mathcal{E}$ and $\mathcal{T}$ are the site-energy and nearest-neighbor 
hopping matrices of the unit cell, respectively. $\mathcal{I}$ is a
($2N_y\times2N_y$) identity matrix. Since in the nano-ribbon translational 
invariance exists along the $x$-direction, we can write $\psi_{j,\sigma}$ 
in terms of the Bloch waves and then Eq.~\ref{diff} takes the form,
\begin{equation}
(E \mathcal{I}-\mathcal{E}_{\sigma})=\mathcal{T} e^{ik_x \Lambda}+
\mathcal{T}^{\dag} e^{-ik_x \Lambda}
\label{bloch}
\end{equation}
where, $\Lambda=\sqrt{3} a$ is the horizontal separation between two
filled magenta or blue circles situated at two successive unit cells.
$a$ is the length of each side of a hexagonal benzene like ring. Solving
Eq.~\ref{bloch} we get the desired energy dispersion relation ($E$ vs.
$k_x$) of the ribbon.

\begin{figure}[ht]
{\centering \resizebox*{7.5cm}{4cm}{\includegraphics{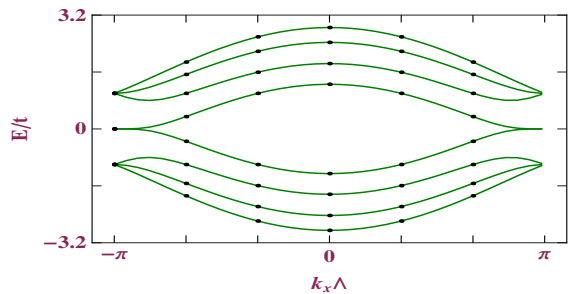}}\par}
\caption{(Color online). Energy levels (green curve) as function of $k_x$ 
for a finite width zigzag nano-ribbon considering $N_y=4$. The discrete 
eigenvalues (filled black circles) of a nanotube with $N_x=12$ and $N_y=4$, 
in the absence of AB flux $\phi$, are superimposed. Here we set $U=0$.}
\label{band}
\end{figure}

As illustrative example, in Fig.~\ref{band} we show the variation of energy 
levels (green curves) as a function of wave vector $k_x$ for a finite width 
zigzag nano-ribbon considering $N_y=4$. Quite interestingly we observe that 
at $E=0$, partly flat bands appear in the spectrum which make the system 
unique. The electronic states corresponding to those almost flat bands are 
characterized by strongly localized states near the zigzag edges of the tube. 
The existence of these edge states have also been reported earlier by some 
other groups~\cite{waka2,brey,neto}. 

Following the energy band structure of a finite width nano-ribbon now
we focus on the variation of energy levels of a nanotube. For a nanotube
$k_x$ also becomes quantized where the quantized values are obtained by
applying periodic boundary condition along the $x$-direction~\cite{waka1}.
The quantized wave numbers are expressed from the relation $k_x=4\pi n_x/
N_x \Lambda$, where $n_x$ is an integer lies within the range:
$-N_x/4\leq n_x< N_x/4$. Plugging the quantized values of $k_x$ in
Eq.~\ref{bloch} we can easily determine the eigenvalues of a finite sized
nanotube. As representative example, in Fig.~\ref{band} we show the 
variation of discrete energy eigenvalues (filled black circles) for a
zigzag nanotube considering $N_x=12$ and $N_y=4$, in the absence of AB flux 
$\phi$ passing through the tube. For this nanotube $k_x$ gets six quantized 
values ($-\pi/\Lambda$, $-2\pi/3\Lambda$, $-\pi/3\Lambda$, $0$, $\pi/3\Lambda$ 
and $2\pi/3\Lambda$), and therefore, total $48$ energy values are obtained 
since $N_y$ is set at $4$.
\vskip 0.2cm
\noindent
{\em \underline{Interacting case}:} In the presence of e-e interaction
energy levels get modified significantly depending on the filling of
\begin{figure}[ht]
{\centering \resizebox*{7.5cm}{7cm}{\includegraphics{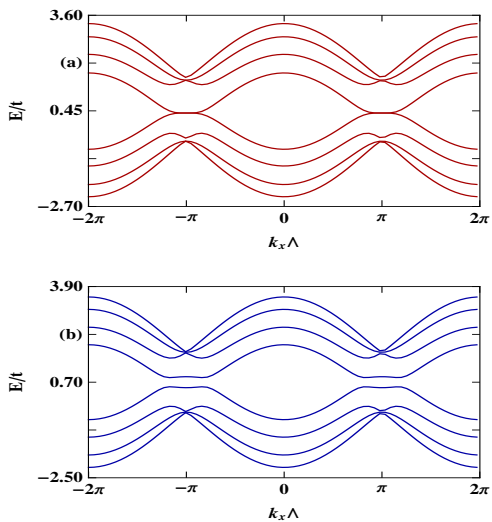}}\par}
\caption{(Color online). Energy levels as function of $k_x$ for a finite 
width zigzag nano-ribbon considering $N_y=4$ and $U=1.4$, where (a) and 
(b) correspond to the one-third- and half-filled cases, respectively.}
\label{uband}
\end{figure}
electrons. The results calculated for a particular value of $U$ are
presented in Fig.~\ref{uband} where we set $N_y=4$. In the half-filled
band case, a gap opens up at the Fermi energy~\cite{rossier} which is
consistent with the DFT calculations~\cite{son} and the gap increases 
with the value of $U$. A careful investigation also predicts that the
full energy band gets shifted by the factor $U/2$.

\section{Second quantized form of persistent current}

In order to evaluate persistent current in individual zigzag paths of a
nanotube, threaded by an AB flux $\phi$, we use second quantized 
approach~\cite{san3,sanu}. This is an elegant and nice way of studying the 
response in separate branches of any quantum network.

We start with the basic equation of current operator 
\mbox{\boldmath$I_{\sigma}$} corresponding to spin $\sigma$ in terms of 
the velocity operator \mbox{\boldmath$v_{\sigma}$}
(=\mbox{\boldmath ${\dot{x}_{\sigma}}$}) as,
\begin{equation}
\mbox{\boldmath $I_{\sigma}$}=-\frac{1}{N_x} e 
\mbox{\boldmath ${\dot{x}_{\sigma}}$}
\label{velo}
\end{equation}
where, \mbox{\boldmath ${x_{\sigma}}$} is the displacement operator. The 
velocity operator is computed from the expression,
\begin{eqnarray}
\mbox{\boldmath $v_{\sigma}$} &=& \frac{1}{i\hbar}
\left[\mbox{\boldmath ${x_{\sigma}}$},\mbox{\boldmath ${H_{\sigma}}$}\right].
\label{equ33}
\end{eqnarray} 
Using this relation we can write the velocity operator of an electron
\begin{figure}[ht]
{\centering \resizebox*{7.5cm}{7cm}{\includegraphics{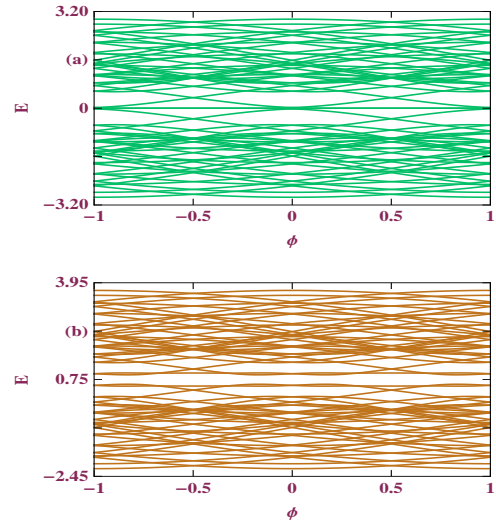}}\par}
\caption{(Color online). Energy-flux characteristics of a half-filled zigzag 
nanotube with $N_x=10$ and $N_y=7$. (a) $U=0$ and (b) $U=1.5$.}
\label{energy}
\end{figure}
with spin $\sigma$ in a zigzag  channel $n$ (say) in the form,
\begin{eqnarray}
\mbox{\boldmath$v_{n,\sigma}$}& = &\frac{t}{i \hbar} \sum_m 
\left(b^{\dag}_{m+1,n,\sigma} a_{m,n,\sigma}e^{-i \theta} \right. \nonumber \\
&-& a^{\dag}_{m,n,\sigma}b_{m+1,n,\sigma}e^{i \theta} -b^{\dag}_{m-1,n,\sigma}
a_{m,n,\sigma}e^{i \theta} \nonumber \\
&+& \left. a^{\dag}_{m,n,\sigma}b_{m-1,n,\sigma}e^{-i \theta} \right).
\label{velop}
\end{eqnarray}
Therefore, for a particular eigenstate $|\psi_{p,\sigma}\rangle$ persistent 
current in $n$-th channel becomes, 
\begin{equation}
I_{n,\sigma}^p=-\frac{e}{N_x}\langle \psi_{p,\sigma}|
\mbox{\boldmath$v_{n,\sigma}$}|\psi_{p,\sigma}\rangle
\label{currequ}
\end{equation}
where, the eigenstate $|\psi_{p,\sigma}\rangle$ looks like,
\begin{eqnarray}
|\psi_{p,\sigma} \rangle &=&\sum_{m,n} \left( \alpha_{m,n,\sigma}^p |
m,n,\sigma\rangle +\beta_{m-1,n,\sigma}^p |m-1,n ,\sigma\rangle \right. 
\nonumber \\
& + & \beta_{m+1,n,\sigma}^p |m+1,n,\sigma \rangle \nonumber \\ 
&+& \left. \beta_{m,n+1,\sigma}^p |m,n+1,\sigma \rangle \right).
\label{shi}
\end{eqnarray}
Here, $|m,n,\sigma\rangle$'s are the Wannier states and 
$\alpha_{m,n,\sigma}^p$ and $\beta_{m,n,\sigma}^p$'s are the corresponding 
coefficients. Simplifying Eq.~\ref{currequ}, we get the final relation of 
persistent charge current for $n$-th zigzag channel as,
\begin{eqnarray}
I_{n,\sigma}^p & = & \frac{iet}{\hbar N_x} \sum_m
\left(\beta^{p~*}_{m+1,n,\sigma} \alpha^p_{m,n,\sigma}e^{-i \theta} 
\right. \nonumber \\
&-&\alpha^{p~*}_{m,n,\sigma}\beta^p_{m+1,n,\sigma}e^{i \theta}
- \beta^{p~*}_{m-1,n,\sigma} \alpha^p_{m,n,\sigma}e^{i \theta} \nonumber \\
&+& \left. \alpha^{p~*}_{m,n,\sigma}\beta^p_{m-1,n,\sigma}e^{-i \theta} 
\right).
\label{vel}
\end{eqnarray} 

Using the same prescription we can also evaluate persistent current in
individual armchair paths (along $y$ direction) of the nanotube. The
final expression of it gets the form,
\begin{widetext}
\begin{eqnarray}
I_{m-1,m,\sigma}^p & = &\frac{i t}{2 \hbar N_y}\left[\sum_{n=1,3,\hdots}^{N_y} 
\left(\alpha^{p~*}_{m,n,\sigma} \beta^p_{m-1,n,\sigma}e^{-i \theta} 
- \beta^{p~*}_{m-1,n},\sigma \alpha^p_{m,n,\sigma}e^{i \theta}
\right) 
+ \sum_{n=2,4,\hdots}^{N_y}\left(\beta^{p~*}_{m,n,\sigma} 
\alpha^p_{m-1,n,\sigma} e^{-i \theta} - \alpha^{p~*}_{m-1,n,\sigma} \right. 
\right. \nonumber \\
& \times &  \beta^p_{m,n,\sigma} e^{i \theta} 
+ \left. \beta^{p~*}_{m-1,n,\sigma} \alpha^p_{m-1,n-1,\sigma}
-\alpha^{p~*}_{m-1,n-1,\sigma} \beta^p_{m-1,n,\sigma} \right)
+ \left. \sum_{n=2,4,\hdots}^{N_y-1} \left( \beta^{p~*}_{m,n+1,\sigma} 
\alpha^p_{m,n,\sigma}-\alpha^{p~*}_{m,n,\sigma} \beta^p_{m,n+1,\sigma} \right)
\right] \nonumber \\
\label{vely}
\end{eqnarray}
\end{widetext}
where, an armchair channel ($m-1,m$) is constructed by ($m-1$)-th and 
$m$-th lines according to our indexing. 

At absolute zero temperature ($T=0\,K$), net persistent current driven by 
electrons of spin $\sigma$ in a 
\begin{figure}[ht]
{\centering \resizebox*{7.5cm}{7cm}{\includegraphics{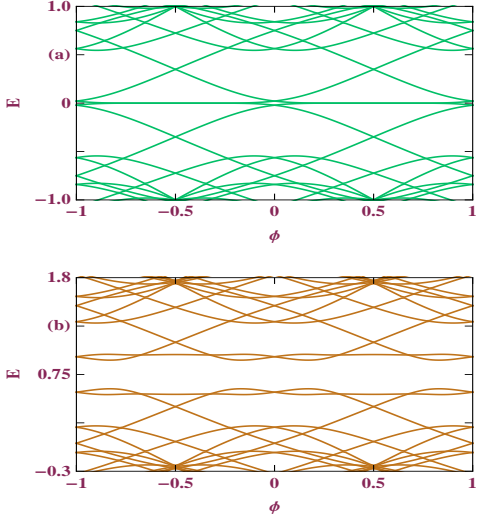}}\par}
\caption{(Color online). Few energy levels of Fig.~\ref{energy} are 
re-plotted for a narrow energy range across the band centres for better 
viewing of the variation of energy levels with flux $\phi$, where (a) and 
(b) correspond to the identical meaning as in Fig.~\ref{energy}.}
\label{spectra2}
\end{figure}
particular channel $n$ for a nanotube described with Fermi energy $E_F$ can 
be determined by taking the sum of individual contributions from the lowest 
energy eigenstates upto the Fermi level. Hence we get,
\begin{equation}
I_{n,\sigma}=\sum_{p} I_{n,\sigma}^p.
\label{pcc}
\end{equation}
Summing $I_{n,\sigma}$ over all possible channels $n$, and $\sigma$ we get 
total persistent current in the nanotube which is mathematically expressed 
as, 
\begin{equation}
I_T=\sum_{n,\sigma} I_{n,\sigma}.
\end{equation}
The persistent current can also be determined in some other ways as 
\begin{figure}[ht]
{\centering \resizebox*{6.7cm}{4cm}{\includegraphics{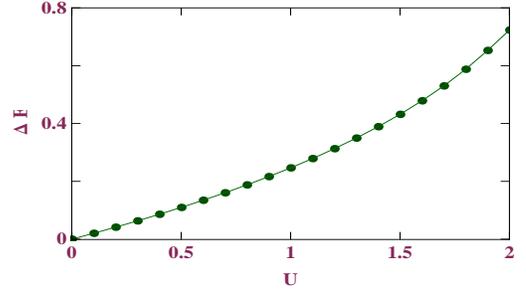}}\par}
\caption{(Color online). Energy gap ($\Delta E$) as a function of on-site 
Hubbard interaction strength $U$ for a zigzag nanotube with $N_x=10$ 
and $N_y=7$ in the half-filled band case when $\phi$ is set at $\phi_0/2$.}
\label{gap}
\end{figure}
available in literature. Probably the simplest way of determining persistent 
current is the case where first order derivative of ground state energy with 
respect to AB flux $\phi$ is taken into account. Therefore, we can write,
\begin{equation}
I_T=-c\frac{\partial E_0(\phi)}{\partial \phi}
\label{deri}
\end{equation}
where, $E_0(\phi)$ is the total ground state energy for a particular electron 
filling. But, in our present scheme, the so-called second quantized approach, 
there are some advantages compared to other available procedures. Firstly, 
we can easily measure current in any branch of a complicated network. 
Secondly, the determination of individual responses in separate branches 
helps us to elucidate the actual mechanism of electron transport in a more 
transparent way.

In the present work we examine all the essential features of persistent
current at absolute zero temperature and use the units where $c=h=e=1$. 
Throughout our numerical calculations we set $t=-1$ and measure all the 
physical quantities in unit of $t$.

\section{Energy-flux characteristics}

In Fig.~\ref{energy} we show the variation of energy levels as a function
of flux $\phi$ for a zigzag nanotube considering $N_x=10$ and $N_y=7$ both
\begin{figure}[ht]
{\centering \resizebox*{8.2cm}{6cm}{\includegraphics{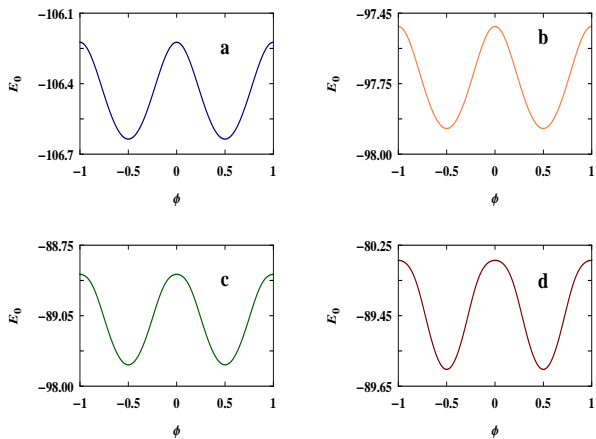}}\par}
\caption{(Color online). Ground state energy level as a function of $\phi$ 
for a zigzag nanotube in the half-filled band case considering $N_x=20$ and 
$N_y=8$. (a), (b), (c) and (d) correspond to $U=0$, $0.5$, $1$ and $1.5$,
respectively.}
\label{ground}
\end{figure}
for the (a) non-interacting and (b) interacting cases. For $U=0$, we 
compute the energy levels simply by diagonalizing the non-interacting 
Hamiltonian and the nature of the energy spectrum becomes independent 
of the total number of electrons $N_e$ in the system. On the other hand, 
for the non-interacting case ($U \ne 0$) we fist decouple the interacting 
Hamiltonian (Eq.~\ref{ham}) for a particular filling, in the mean field 
scheme, into two non-interacting Hamiltonians (for up and down spin 
electrons) and then diagonalize the Hamiltonian for up (down) spin electrons. 
For identical filling factor of up and down spin electrons the energy levels 
are exactly similar both for $H_{\uparrow}$ and $H_{\downarrow}$ (see 
Fig.~\ref{energy}(b)), and therefore, one energy spectrum cannot be 
separated from the other. Since in our case we set $N_x=10$ and $N_y=7$, 
we get total $70$ independent energy levels and due to their overlaps 
individual energy levels are not clearly distinguished from the spectra 
given in Fig.~\ref{energy}. To have a better look in Fig.~\ref{spectra2} 
we re-plot few energy levels of Fig.~\ref{energy} collecting them from a 
narrow energy range across the band centres, where (a) and (b) correspond 
to the identical meaning as in Fig.~\ref{energy}. From the spectra we
see that all the energy levels vary periodically with $\phi$ providing 
$\phi_0$ ($=1$ in our chosen unit) flux-quantum periodicity. At half-integer 
or integer multiples of $\phi_0$, energy levels have either a maximum or a 
minimum (see Fig.~\ref{spectra2}), and accordingly, at these points 
persistent current becomes zero which is quite obvious since the
current is obtained by taking the first order derivative of the eigenenergy 
with respect to flux $\phi$ (Eq.~\ref{deri}). Both the energy spectra take a 
complicated look as there are many crossings among the energy levels 
particularly in the regions away from $E=0$ (Fig.~\ref{spectra2}). At $E=0$, 
the energy levels become almost flat for a wide range of $\phi$, and, near 
$\phi=\pm \phi_0/2$ they vary slowly with $\phi$ as shown in 
Fig.~\ref{spectra2}(a). These almost flat energy levels support a very 
\begin{figure}[ht]
{\centering \resizebox*{8.2cm}{6cm}{\includegraphics{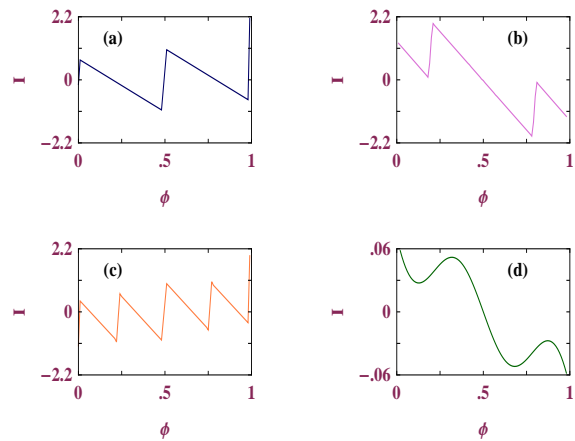}}\par}
\caption{(Color online). Current-flux characteristics of a zigzag nanotube 
with $N_x=14$ and $N_y=6$, where (a), (b), (c) and (d) correspond to 
$N_e=10$, $20$, $30$ and $82$, respectively. $U$ is fixed at $1$.}
\label{filling}
\end{figure}
little contribution to the persistent current as clearly followed from 
Eq.~\ref{deri}. While, the other energy levels with larger slopes provide 
large persistent current. This peculiar nature of the energy levels invokes 
the {\em current amplitude to become filling dependent} and we elaborate it 
in the following section. In Fig.~\ref{energy}(b) we display the variation 
of energy levels with $\phi$ for a zigzag nanotube with the same parameter 
values declared above in the presence of Hubbard interaction. Here we choose 
$U=1.5$. Both for the up and down spin Hamiltonians the eigenvalues are 
exactly identical and they overlap with each other. Few energy levels of 
Fig.~\ref{energy}(b) are also re-plotted in Fig.~\ref{spectra2}(b) for better 
viewing. The electronic correlation leads to an energy gap at the band centre 
and the gap increases with $U$. It is illustrated in Fig.~\ref{gap}. This 
energy gap is consistent with the energy gap obtained in the $E$-$k_x$ 
diagram (Fig.~\ref{uband}(b)).

The variation of ground state energy level of a carbon nanotube with
zigzag edges as a function of magnetic flux $\phi$ is depicted in 
\begin{figure*}[ht]
{\centering \resizebox*{17cm}{12cm}{\includegraphics{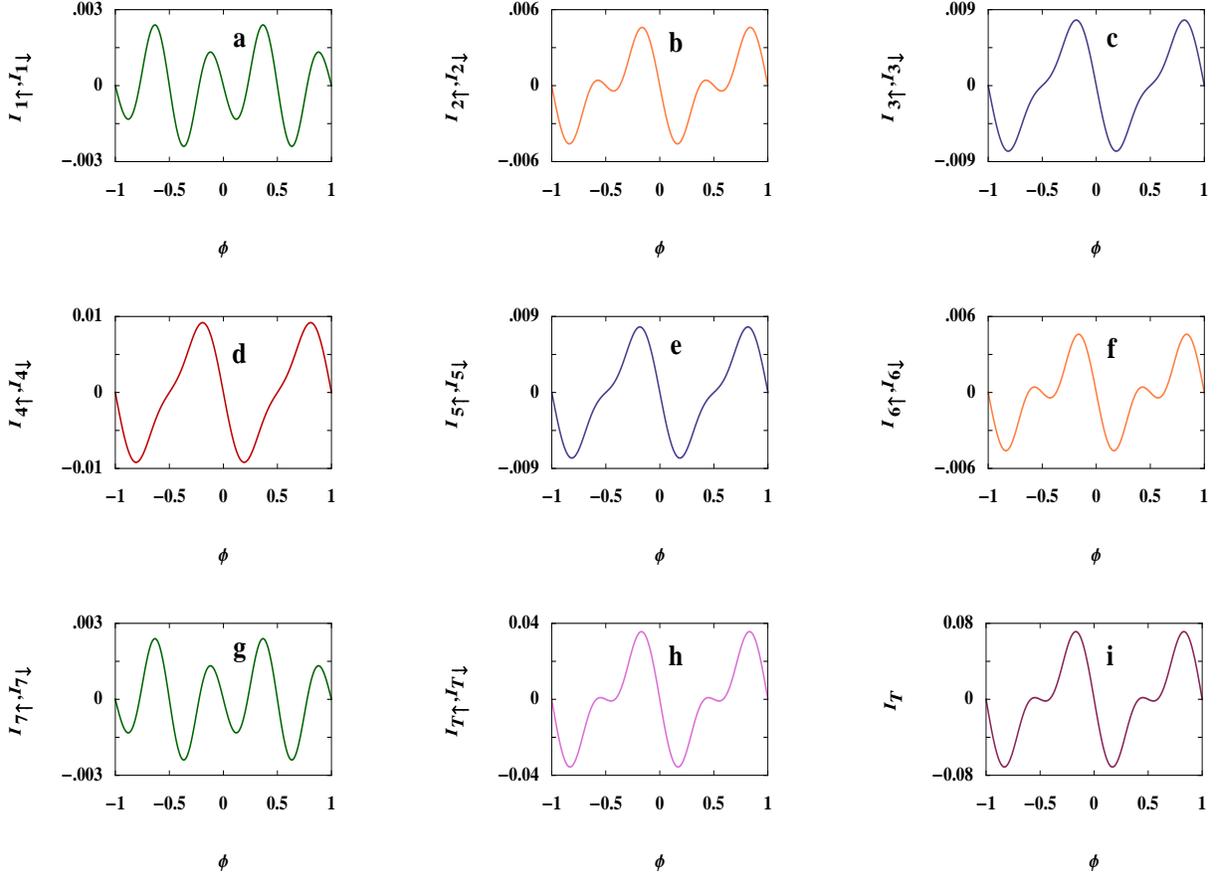}}\par}
\caption{(Color online). Persistent current in individual zigzag paths as a
function of $\phi$ for a half-filled zigzag nanotube ($N_x=20$ and $N_y=7$)
with $U=1.2$, where, (a)-(g) correspond to $1$st-$7$th zigzag channels of 
the tube, respectively. The net current corresponding to both up and down 
spin electrons are displayed in (h) while in (i) total persistent current is 
shown.}
\label{udistribution}
\end{figure*}
Fig.~\ref{ground} in the half-filled band case, where (a), (b), (c) and 
(d) correspond to the four different values of electronic correlation
strength $U=0$, $0.5$, $1$ and $1.5$, respectively. The energy levels 
evince one flux-quantum periodicity, as expected, and their energies get
increase with $U$. It can be explained as follows. In presence of $U$, 
each site is occupied by an electron with either up or down spin in the 
half-filled band case. This is the ground state energy configuration. 
Now, if we add an electron further, the probability of getting two opposite
spin electron in a single site becomes finite which gives higher energy
due to the repulsion in presence of $U$.

\section{Current-flux characteristics}

Now we focus our attention on the behavior of persistent current in a 
zigzag nanotube. 

First, we illustrate the dependence of persistent current amplitude on
the electron filling. To ensure it in Fig.~\ref{filling} we present the 
current-flux characteristics of a zigzag nanotube considering $N_x=14$ and
$N_y=6$, where four different figures correspond to the four different cases
of electron filling. The Hubbard interaction strength is set at $1$. It is
observed that when the number of electrons is much 
smaller than half-filling, persistent current exhibits multiple kinks at 
different values of $\phi$, associated with the multiple crossings of 
energy levels, as shown in Figs.~\ref{filling}(a)-(c). In these three cases
\begin{figure}[ht]
{\centering \resizebox*{6.5cm}{4cm}{\includegraphics{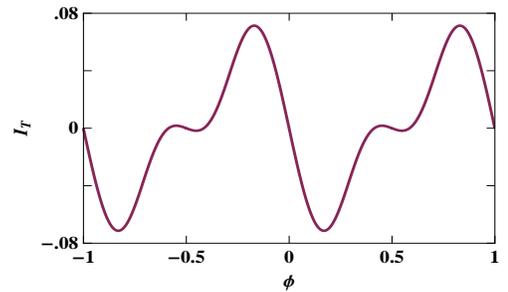}}\par}
\caption{(Color online). Total persistent current obtained in a traditional 
derivative approach (Eq.~\ref{deri}) as a function of $\phi$ for the same 
parameter values mentioned in Fig.~\ref{udistribution}.}
\label{total}
\end{figure}
the number of electrons are $10$, $20$ and $30$, respectively. This is 
quite analogous to the nature of persistent current observed in 
conventional multi-channel mesoscopic cylinders. The behavior of persistent 
current gets significantly modified when the nanotube becomes half-filled 
or nearly half-filled. For example, see Fig.~\ref{filling}(d). Here we
choose $N_e=82$ i.e., the nanotube is very near to the half-filled band case.
In such a case all the kinks disappear and current varies almost continuously,
analogous to the behavior of persistent current observed in traditional 
single-channel mesoscopic rings. For the cases when the nanotube is far away
from half-filling, current amplitudes are quite comparable to each other 
(see Figs.~\ref{filling}(a)-(c)). On the other hand, when the tube is
nearly half-filled current amplitude remarkably gets suppressed. It is 
\begin{figure}[ht]
{\centering \resizebox*{6.5cm}{4cm}{\includegraphics{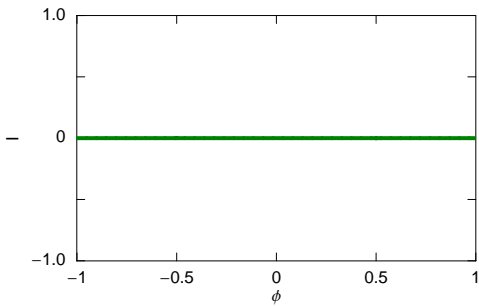}}\par}
\caption{(Color online). Persistent current in an armchair path as a 
function of $\phi$ for a zigzag nanotube with the same parameter values as 
mentioned in Fig.~\ref{udistribution}.}
\label{yvelocity}
\end{figure}
illustrated in Fig.~\ref{filling}(d). This enormous reduction of current 
amplitude can be visualized from the $E$-$\phi$ characteristics given in 
Fig.~\ref{energy}. At half-filling or very close to half-filling, 
the top most filled energy level lies in the nearly flat region i.e.,
around $E=0$ (see Fig.~\ref{energy}(a)) and it contributes a little to 
the current. Moreover, when $U\ne0$, there is gap in the midband region. 
Now, for a particular filling the net persistent current is obtained by 
taking the sum of individual contributions from the lowest filled energy 
levels, and, in this process only the contribution which comes from the 
highest occupied energy level survives finally and the rest disappear due 
to their mutual cancellations. It leads to the enormous reduction of 
persistent current amplitude in the half-filled or nearly half-filled case. 
This feature is independent of the size of the nanotube. In all these cases, 
persistent current varies periodically with flux $\phi$, exhibiting $\phi_0$ 
flux-quantum periodicity. From these current-flux spectra we can emphasize 
that, the current amplitude in a zigzag nanotube is highly sensitive to the 
electron filling and this phenomenon can be utilized {\em in designing a high 
conducting to a low conducting switching operation and vice versa.}

Finally, we concentrate on the behavior of persistent current in 
separate branches of a zigzag carbon nanotube. As illustrative examples, 
in Fig.~\ref{udistribution} we show the variation of persistent current in 
individual zigzag paths as a function of flux $\phi$ for the half-filled 
case considering $N_x=20$ and  $N_y=7$, where (a)-(g) correspond to 
$1$st-$7$th zigzag channels of the tube, respectively. The Hubbard
correlation strength $U$ is set equal to $1.2$. In each of these figures 
we display currents carried by both up and down spin electrons together. 
They are exactly superposed with each other. All these 
currents exhibit $\phi_0$ flux-quantum periodicity and their magnitudes
are quite comparable to each other. Interestingly we see that 
$I_{1\uparrow}$ is exactly identical to $I_{7\uparrow}$, and, similarly 
for the ($I_{2\uparrow}$, $I_{6\uparrow}$) and ($I_{3\uparrow}$, 
$I_{5\uparrow}$) pairs. $I_{4,\uparrow}$, the current in the middle channel, 
becomes the isolated one since we have chosen $N_y=7$. This is true for any 
zigzag nanotube with odd $N_y$. For a tube with even $N_y$, currents are 
pairwise identical. Summing up the individual currents in seven zigzag 
channels we get net persistent current carried by up and down spin 
electrons in the nanotube which is presented in Fig.~\ref{udistribution}(h) 
and the total current is displayed in Fig.~\ref{udistribution}(i) which 
exactly matches with the total current derived from the conventional 
method where first order derivative of the ground state energy is taken 
into account, as shown in Fig.~\ref{total}. It emphasizes that the net 
contribution of persistent current in a zigzag carbon nanotube comes only 
from the individual zigzag channels, not from the armchair paths. To justify 
it in Fig.~\ref{yvelocity} we present the variation of persistent current 
in an armchair path as a function of $\phi$ for a half-filled zigzag nanotube 
considering $N_x=20$ and $N_y=7$, which clearly shows zero current for the 
entire range of $\phi$. 

\section{Summary}

To conclude, in the present work we investigate in detail the magnetic 
response of a zigzag nanotube, threaded by a magnetic flux $\phi$, using 
a generalized Hartree-Fock mean field approach. The model is described by
a simple tight-binding framework. Following the $E$-$k_x$ spectra of both
the non-interacting and interacting cases of a finite width nanoribbon,
we present the results of a nanotube with zigzag edges. Energy levels get
modified significantly in the presence of Hubbard interaction and the 
nature of the energy spectrum strongly depends on the electron filling.
At the half-filled bad case, a gap opens up at the Fermi energy which
is consistent with the DFT calculations. Next, we establish the 
second quantized form to evaluate persistent current in individual
paths of a zigzag carbon nanotube. From the current-flux characteristics 
we can emphasize that the current amplitude in the zigzag nanotube is
highly sensitive to the electron filling and this phenomenon can be utilized
in designing a high conducting to a low conducting switching device and
vice versa.
 
\vskip 0.3in
\noindent
\begin{center}
{\bf\small ACKNOWLEDGMENTS}
\end{center}
\vskip 0.2in
\noindent
First author (PD) thanks M. Dey and N. S. Das for their help in drawing the 
colored figures of the nanoribbon and tube.

\end{document}